\documentclass[page-classic]{epl2}
\usepackage{latexsym}
\usepackage{amssymb}
\usepackage{amsmath}

\begin{document}

\title{Variable enstrophy flux and energy spectrum in two-dimensional turbulence with Ekman friction}

\author{Mahendra K. Verma\inst{1} }

\institute{                    
  \inst{1} Department of Physics, Indian Institute of Technology, Kanpur, India 208016
}


\date{\today}

\abstract{ Experiments and numerical simulations reveal that in the forward cascade regime, the energy spectrum of two-dimensional turbulence with Ekman friction deviates from  Kraichnan's prediction of $k^{-3}$ power spectrum.  In this letter we explain this observation using an analytic model based on variable enstrophy flux arising due to Ekman friction.  We derive an expression for the enstrophy flux which exhibits a logarithmic dependence in the inertial range for the Ekman-friction dominated flows.  The energy spectrum obtained using this enstrophy  flux shows a power law scaling  for large Reynolds number and small Ekman friction, but has an exponential behaviour for large Ekman friction and relatively small Reynolds number.  }

\maketitle
Physics of turbulent flow is quite complex. One of the important and generic features of turbulent flow in three dimensions (3D) is a constant energy flux from large length scales to small length scales.  The wavenumbers exhibiting constant energy flux have $k^{-5/3}$ energy spectrum~\cite{Pope:book}.   Two-dimensional (2D) fluid turbulence however has significantly different behavior.   Kraichnan~\cite{Kraichnan:1971JFMb} showed that in 2D turbulence, the low wavenumber modes (below the forcing wavenumber) exhibit inverse energy cascade, while the large wavenumber modes have forward enstrophy  (square of the vertical vorticity) cascade.  These two regimes have $k^{-5/3}$ and $k^{-3}$ energy spectra respectively.  The above features have been observed in numerical simulations~\cite{Siggia:1981PF,Frisch:1984PF,Boffetta:2007JFM,Tabeling:2002PR}, and in experiments involving electromagnetically driven flows~\cite{Sommeria:1986JFM,Paret:1997PRL} and  soap films~\cite{Rutgers:1998PRL,Vorobieff:1999PF}.   The atmospheric data  indicates that some features of the atmospheric turbulence is two-dimensional~\cite{Boer:1983JAS,Lindborg:1999JFM}. 

The fluid flow in a two dimensional surface is also affected by the drag from its environment  as demonstrated by several experiments~\cite{Belmonte:1999PF,Rivera:2000PRL,Rivera:2001PRL,Kellay:2002RPP,Boffetta:2007EPL} and numerical simulations~\cite{Nam:2000PRL,Boffetta:2002PRE}.   This kind of friction is also referred to as ``Ekman friction".  Belmonte et al.~\cite{Belmonte:1999PF}  performed experiments on freely suspended soap films, and reported that the energy spectrum is steeper than $k^{-3}$.   Boffetta et al.~\cite{Boffetta:2007EPL} observed similar steepening of the spectrum in their experiment on a thin layer of ionic fluids, which is driven electromagnetically and is moving under a layer of fresh water.  Here, the drag on the ionic fluid is induced by the shear layer induced by fresh water~\cite{Rivera:2000PRL,Boffetta:2007EPL}.   Nam {\em et al.}~\cite{Nam:2000PRL} and Boffetta {\em et al.}~\cite{Boffetta:2002PRE} performed numerical simulations with Ekman friction and reported the spectral exponent to be larger than 3.  

The aforementioned steepening of the energy spectrum in two-dimensional flow has been attributed to Ekman friction, which is modeled as $-\alpha {\bf u}$, where $\alpha$ is a constant, and ${\bf u}$ is the velocity field.   Nam {\em et al.}~\cite{Nam:2000PRL} and Boffetta et al.~\cite{Boffetta:2007EPL} attempted to derive the new spectral indices using an analogy with the dynamics of a scalar in a turbulent fluid.  They postulated that a fluid blob has a finite lifetime $\tau = \alpha^{-1}$, and it is passively advected.   The turbulent motion leads to a stretch of a fluid blob with a mean rate given by the Lyapunov exponent $\lambda$. Using the above assumption, the incompressibility condition, and  the mean field approximation, Boffetta et al.~\cite{Boffetta:2007EPL} predicted that the energy spectrum $E(k) \sim k^{-3-2\alpha/\lambda}$.  Boffetta {\em et al.}~\cite{Boffetta:2002PRE} also studied the intermittency effects of Ekman friction, and showed that the small-scale statistics of the vorticity fluctuations is related to the passive scalar transported by the velocity field.  Perleker and Pandit~\cite{Perlekar:2009NJP} performed detailed numerical simulation and studied the effects of Ekman friction on the structure function.  They showed that the velocity structure functions display simple scaling. The reader is also referred to a review article by Kellay and Goldburg~\cite{Kellay:2002RPP}.

The arguments of Boffetta et al.~\cite{Boffetta:2007EPL} involves several assumptions and relatively complex computation of the Lyapunov exponent.  In the present letter we compute the energy spectrum using a model based on a variable enstrophy flux.  Unlike the viscous force, which affects the dissipation range,  Ekman friction is active at all scales.  As a result,  the enstrophy flux decreases significantly in the inertial range itself.  This decrease in the flux leads to a steepening of the energy and enstrophy spectra.    In the present letter, we will derive an expression for the variable enstrophy flux in terms of the dissipative parameters $\alpha$, the kinematic viscosity $\nu$, and other turbulence parameters.  We will show that the variable enstrophy flux yields energy spectrum steeper than $k^{-3}$.

The dynamical equations for the vertical vorticity in a two-dimensional fluid flow with Ekman friction are
\begin{equation}
\frac{\partial \omega} {\partial t} + {\bf u \cdot} \nabla \omega  =   
	-\alpha \omega +  \nu \nabla^2 \omega + f(t),
	\label{eq:omegax}
\end{equation}
and the incompressibility constraint $\nabla \cdot {\bf u} = 0$, which implies that the density of the fluid is a constant.  In the above equation, ${\bf u}$ is the velocity field, $\omega = (\nabla \times {\bf u})_z$ is the vertical component of the vorticity field,  and $\nu$ is the kinematic viscosity of the fluid.  An external force $f(t)$ is applied to maintain a steady-state.   

The  enstrophy flux $\Pi(k)$ is defined as the total enstrophy transferred from the modes inside the wavenumber sphere of radius $k$ to the modes outside the sphere.  Since the above flux is defined in the wavenumber space, we rewrite Eq.~(\ref{eq:omegax}) in the Fourier space as
\begin{equation}
\frac{\partial \omega ({\bf k)}} {\partial t}   = - i k_j \sum_{\bf q} u_j ({\bf q}) \omega ( {\bf k-q})
	-\alpha \omega({\bf k}) - \nu k^2  \omega({\bf k}) + f({\bf k}),
\end{equation} 
where  ${\bf k}$ is the wavenumber.   The corresponding enstrophy evolution equation is
\begin{equation}
\frac{\partial Z(k)} {\partial t}   = T(k) -2\alpha Z(k) - 2\nu k^2  Z(k) + F(k),
\end{equation} 
where  $Z(k)= |\omega({\bf k})|^2/2$ is the one-dimensional enstrophy spectrum, $T(k)$ is the enstrophy transfer term arising due to nonlinearity, and $F(k)$ is the enstrophy supply rate due to the forcing~\cite{Pope:book}.   It has been analytically shown that the total energy $E=\sum |{\bf u(k)}|^2/2$ and the total enstrophy $Z = \sum |\omega({\bf k})|^2/2$ are conserved in the inviscid limit ($\nu=0$ and $\alpha=0$) without any external forcing~\cite{Kraichnan:1971JFMb,Pope:book}.  The dual energy spectrum described in the introduction is a consequence of simultaneous conservation of energy and enstrophy~\cite{Kraichnan:1971JFMb}.  Two-dimensional turbulent flow with $\alpha = 0$ and $\nu \rightarrow 0$ exhibits $k^{-5/3}$ energy spectrum for $k<k_f$, and $k^{-3}$ energy spectrum for $k>k_f$, where $k_f$ is the center of the forcing-wavenumber band.  The scenario changes significantly when $\alpha \ne 0$, as observed in earlier experiments and numerical simulations.

The forcing wavenumber in a typical experiment on Ekman friction is of the order of the box size.  Consequently, experiments report only forward enstrophy cascade and energy spectrum as $k^{-a}$  with $a \ge 3$.   Since $|\omega(k)| = k |{\bf u(k)}| $, the enstrophy spectrum would vary as $k^{-a+2}$.   Pope~\cite{Pope:book} prescribed a function for the energy spectrum for three-dimensional fluid turbulence.  We adapt Pope's function for the enstrophy spectrum $Z(k)$ as
\begin{equation}
Z(k) = C (\Pi(k))^{2/3} k^{-1}  f_L(k L) f_\eta(k \eta),
\label{eq:Zk} 
\end{equation}
where $C\approx 1.4$ is a constant (equivalent to Kolmogorov's constant for 3D fluid turbulence)~\cite{Lesieur:book,Lindborg:2010PF}, $\Pi(k)$ is the enstrophy flux emanating  from the wavenumber sphere of radius $k$, and $f_L(kL)$,$f_\eta(k \eta)$ specify the components of the  forcing-scale and dissipative-scale enstrophy spectra respectively.   These functions have been described by Pope~\cite{Pope:book} as
\begin{eqnarray}
f_L(kL) & = & \left( \frac{kL}{[(kL)^2 + c_L]^{1/2}} \right)^{1+p_0},  \\
f_\eta(k\eta) & = & \exp \left[ -\beta \left\{ [ (k\eta)^4 + c_\eta^4 ]^{1/4}   - c_\eta \right\} \right],
\end{eqnarray}
where $L$ is the box size, and $c_L, c_\eta, p_0, \beta$ are constants, which are determined by matching the function of Eq.~(\ref{eq:Zk}) with experimental observations.  In the absence of any clear prescription for these constants, we take  $C_L \approx 6.78$, $c_\eta \approx 0.40$, $\beta \approx 5.2$ and $p_0 =2$, used by Pope~\cite{Pope:book} for 3D turbulence.  

In Kraichnan's phenomenology for the two-dimensional fluid turbulence without Ekman friction, the enstrophy flux $\Pi(k)$ is a constant in the inertial range.  This is due to the fact that in this regime, the local dissipation rate ($2 \nu k^2 Z(k)$) is negligible, and the forcing is absent.  However, in the presence of Ekman friction, $\Pi(k)$ decreases in the inertial range itself due to the $ 2 \alpha Z(k)$ term.  The enstrophy fluxes $\Pi(k+dk)$ differs from $\Pi(k)$ by the rate of enstrophy loss in the wavenumber shell $(k,k+dk)$, i.e., 
 \begin{equation}
 \Pi(k+dk) - \Pi(k) = - \left\{2\nu k^2 + 2\alpha \right\} Z(k) dk,
\end{equation}
or
 \begin{equation}
 \frac{d\Pi(k)}{dk} = - \left\{2\nu k^2 +2 \alpha \right\} C (\Pi(k))^{2/3} k^{-1} f_\eta(k \eta).
 \label{eq:dPidk}
\end{equation}
In the present letter, we focus on the inertial and dissipative ranges.  Therefore,  we have ignored the variation of the forcing-scale spectrum $f_L(kL)$, i.e., $f_L(kL) \approx 1$.

Equation~(\ref{eq:dPidk}) can be easily integrated, which yields
\begin{eqnarray}
\left[ \frac{\Pi(k)}{\Pi_0} \right]^{1/3} & = & 1- \frac{2C}{3}  \frac{\nu}{\Pi_0^{1/3} \eta^2}  I_1(k \eta)
			- \frac{2\alpha C}{3 \Pi_0^{1/3}} I_2(k\eta) \nonumber \\
		& = & 1-\frac{2C }{3} C_1   I_1(k \eta) - \frac{2\alpha C}{3 \Pi_0^{1/3}}  I_2(k\eta),
		\label{eq:Pi_analytic}
\end{eqnarray}
where $\Pi_0$ is the maximum value of  the enstrophy flux at $k=k_1$ (which is also the lower limit for the integration), $C_1 = \nu/(\Pi_0^{1/3} \eta^2)$ is a constant, and
\begin{equation}
\eta = \frac{1}{\sqrt{C_1}} \frac{\sqrt{\nu}}{\Pi_0^{1/6}}
\label{eq:eta}
\end{equation}
is the Kolmogorov length for two-dimensional turbulence.  For all our calculations, we also take the constant $C$ to be 1.4 ~\cite{Lesieur:book,Lindborg:2010PF}.    The integrals $I_1$ and $I_2$ are
\begin{eqnarray}
I_1(k \eta) & = & \int_{k_1 \eta}^{k \eta} dk' k' f_\eta(k') \\
I_2(k \eta) & = & \int_{k_1 \eta}^{k \eta} dk' k'^{-1} f_\eta(k') 
\end{eqnarray}
where the lower limit for the wavenumber $k_1$ is chosen to be $ 6 \times 2 \pi/L$~\cite{Pope:book}.  Pope~\cite{Pope:book} suggests that the inertial range starts around this wavenumber.  Using Eq.~(\ref{eq:eta}) and  $\Pi_0 = U^3/L^3$, we can deduce that $k_1 \eta = 6 (2 \pi/L) \eta =12\pi /\sqrt{C_1 Re}$.   In our model the enstrophy flux peaks at $k=k_1$ and  decreases for higher values of $k$.   Equation~(\ref{eq:Pi_analytic}) shows that the enstrophy flux depends critically on the nondimensional parameter  $\alpha' =2 \alpha C /(3 \Pi_0^{1/3})$.   

The form of Eq.~(\ref{eq:Pi_analytic}) appears quite complex, but it can easily approximated in the inertial range in the $\nu \rightarrow 0$  limit.  Under this limit,  $I_1(k) \approx 0$ and  $I_2(k) \approx \alpha' \log(k/k_1)$.   Therefore, the approximate enstrophy flux would be
\begin{equation}
\frac{\Pi(k)}{\Pi_0}  \approx  \left[ 1 - \alpha' \log(k/k_1) \right]^{3},
\label{eq:Pi_approximate}
\end{equation}
which shows that for nonzero $\alpha'$, the enstrophy flux  decreases logarithmically with the increase of wavenumber.  We will show later that  for turbulent flows, the above approximate function matches quite well with that computed using Eq.~(\ref{eq:Pi_analytic}) in the inertial range. Additionally, the wavenumbers satisfying the condition $\alpha' \log(k/k_1) \ll 1$ follow a power law behavior:
\begin{equation}
\frac{\Pi(k)}{\Pi_0}  \approx  e^{-3 \alpha' \log(k/k_1)} \approx (k/k_1)^{-3\alpha'}
\label{eq:Pi_powerlaw}
\end{equation}

Once we have computed the $k$-dependent enstrophy flux, the enstrophy spectrum can be immediately derived as
\begin{equation}
Z(k) =
\begin{cases} 
C \Pi_0^{2/3} k^{-1}   f_\eta(k \eta) \left[ \frac{\Pi(k)}{\Pi_0} \right]^{2/3}, & \text{if  $k>k_1$}  \\
C \Pi_0^{2/3} k^{-1}  f_L(k L), & \text{otherwise,}
\end{cases}
\end{equation}
while the energy spectrum varies as
\begin{equation}
E(k) =
\begin{cases} 
C \Pi_0^{2/3} k^{-3}   f_\eta(k \eta) \left[ \frac{\Pi(k)}{\Pi_0} \right]^{2/3}, & \text{if  $k>k_1$}  \\
C \Pi_0^{2/3} k^{-3}  f_L(k L), & \text{otherwise.}
\end{cases}
\label{eq:Ek}
\end{equation}
For a narrow range of the inertial range with $\alpha' \log(k/k_1) \ll 1$, the aforementioned equations yield
 \begin{equation}
E(k)  \sim  C \Pi_0^{2/3} k^{-3-2\alpha'}.
\label{eq:Ek_3plus2alpha}
\end{equation}

\begin{figure}[htbp]
 \begin{center}
 \includegraphics[scale=0.32]{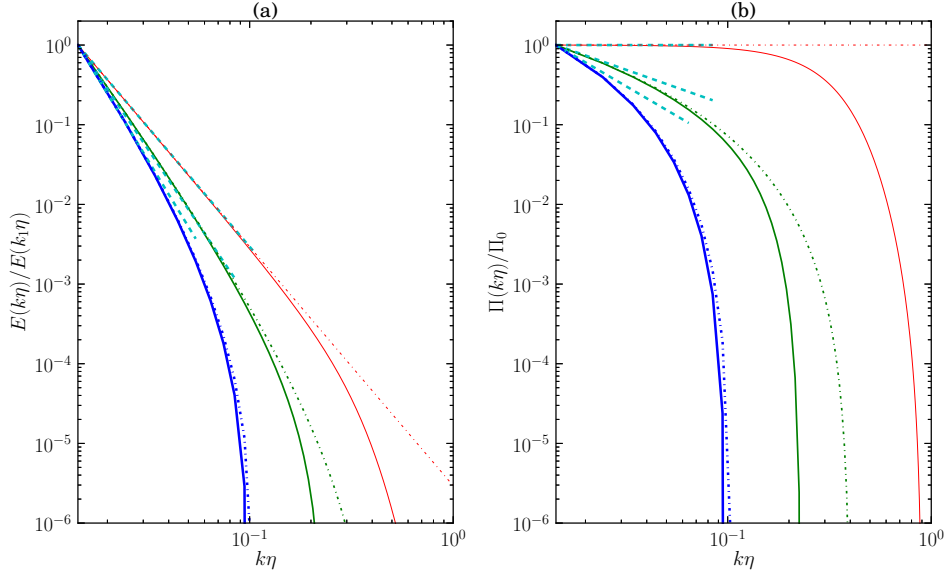}
 \end{center}
\caption{  (a) Plots of energy spectra for $Re=10^6$ and $\alpha'=0$ (red thin curve), 0.3 (green thicker curve), and 0.5 (blue thickest curve).  The best fit curves $k^{-a}$ with $a=3, 3.8$ and 4.2 are plotted as dashed cyan lines.  The approximate energy spectrum using the approximate enstrophy flux of Eq.~(\ref{eq:Pi_approximate}) is exhibited as chained lines.  (b) Plots of the corresponding enstrophy flux (solid lines), the approximate enstrophy flux of Eq.~(\ref{eq:Pi_approximate}) (chained lines), and its power law approximation described by Eq.~(\ref{eq:Pi_powerlaw}) (cyan dashed lines). }  
 \label{fig:Re1e6}
 \end{figure}
 
We can also compute the enstrophy dissipation rates due to the viscous force and Ekman friction using $D_\nu(k) =  \int_0^\infty 2 \nu k^2 Z(k) dk$ and $D_\alpha(k)  =  \int_0^\infty 2 \alpha  Z(k) dk$ respectively.  Another important quantity is the ratio of the nonlinear term and Ekman friction, which can be estimated using $\operatorname{Re}_\alpha = U/(\alpha L)$, termed as ``Reynolds number based on Ekman friction".   We will show later that $\operatorname{Re}_\alpha$ plays an important role in determining the structure of the energy spectrum.

We use the aforementioned model to study the nature of energy spectrum as a function of the Ekman friction parameter $\alpha$.  We take two cases: large Reynolds number, and relatively small Reynolds number.   For the former case, we take $\operatorname{Re}=10^6$ and $\alpha'=0,0.3$, and 0.5, and  compute the normalized enstrophy flux $\Pi(k\eta)/\Pi_0$  and energy spectrum $E(k\eta)/E(k_1\eta)$ using equations (\ref{eq:Pi_analytic}) and (\ref{eq:Ek}) respectively.  These functions are exhibited in Fig.~\ref{fig:Re1e6} using red ($\alpha'=0$), green ($\alpha'=0.3$), and blue ($\alpha'=0.5$) curves.  For $\alpha'=0$, the inertial range enstrophy flux is nearly constant and $E(k) \sim k^{-3}$, consistent with Kraichnan's predictions, more so a validation of our model.   Note however that $\Pi(k)$ must vanish for wavenumbers $k \gg \eta^{-1}$.  We use this condition to determine the constant $C_1$ of Eq.~(\ref{eq:Pi_analytic}), which yields $C_1 = 7.17$.

With the increase of $\alpha'$ or Ekman friction, the entropy flux becomes $k$-dependent and the energy spectrum is steeper than $k^{-3}$.  For $\alpha'=0.3$ and 0.5, the spectral indices are 3.8 and 4.2 respectively.   The best fit curves are shown in Fig.~\ref{fig:Re1e6}(a) using the dashed cyan lines.  We also plot the approximate enstrophy flux of Eq.~(\ref{eq:Pi_approximate}) and the corresponding energy spectrum in Fig.~\ref{fig:Re1e6} as chained lines.  In the inertial range, the approximate energy spectrum and the approximate enstrophy flux fit rather well with the model predictions.    For a very narrow range of $k\eta$, $\Pi(k) \sim k^{-3\alpha'}$ as shown by the cyan dashed line in Fig.~\ref{fig:Re1e6}(b).  This result indicates that the the logarithmic dependence of enstrophy flux is valid for a larger range of wavenumbers than the power law of Eq.~(\ref{eq:Pi_powerlaw}). This observation is also consistent with the fact that the spectral index of the energy spectrum is not exactly equal to $3+2\alpha'$ [Eq.~(\ref{eq:Ek_3plus2alpha})], but these two exponents are reasonably close.  

We also compute $\operatorname{Re}_\alpha$ for $\alpha'=0.3$ and 0.5 and find them to be approximately 3 and 2 respectively (see Table 1).  This result indicates that the nonlinear term ${\bf u \cdot \nabla u}$ dominates  Ekman friction, consistent with the power law behaviour of  the energy spectrum.  The ratio $D_\alpha/D_\nu$ is 22 and 100 for $\alpha'=0.3$ and 0.5 respectively, indicating that the Ekman friction is stronger than the viscous term.  The strong wavenumber dependence of the enstrophy flux is due to the important role played by Ekman friction.  
 
 We perform our model calculations for the experimental parameters of  Boffetta et al.~\cite{Boffetta:2007EPL}.  The Reynolds number for these runs ranges from 3950 to 6600, which is relatively small compared to the aforementioned test cases with $\operatorname{Re} = 10^6$.  In Boffetta et al.'s experiment,  a thin layer of electrolyte solution of water and NaCl contained in a square tank ($L=$ 50 cm) is stirred electromagnetically using four triangular shaped electromagnets with alternating signs.   A layer of fresh water was kept above the electrolyte solution.  The thickness of the ionic fluid was maintained at 0.3 cm, while that of the upper fluid was varied from 0.3 cm to 0.7 cm.   The ionic fluid experiences a drag force $-\alpha {\bf u}$.   The kinematic viscosity of the electrolyte is same as that of water ($\nu \approx 0.01$ $\mathrm{cm^2/sec}$).  For the three sets of experiments performed,  Boffetta et al.~reported the Ekman friction coefficient $\alpha = 0.037,0.059,0.069$, $\operatorname{Re} = u_\mathrm{rms} L /\nu = 6600,6650,3950$, and the spectral indices $a= 3.5, 3.8$, and 4.0 respectively.  See Table 1 for more details.

 \begin{figure}[htbp]
 \begin{center}
 \includegraphics[scale=0.25]{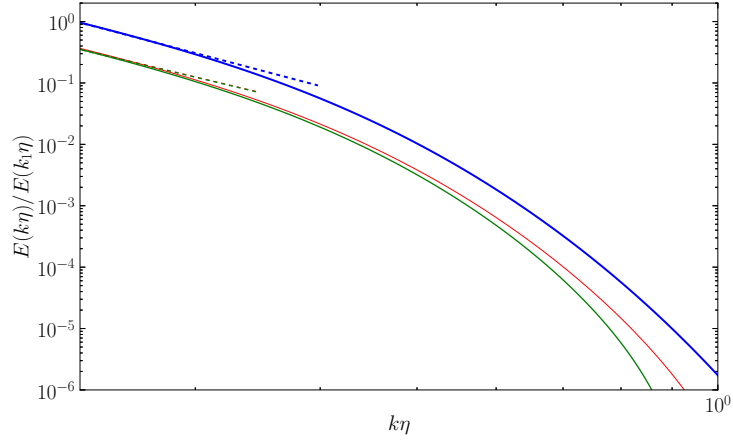}
 \end{center}
 \caption{ Plots of the energy spectra $E^u(k\eta)/E^u(k_1 \eta)$ computed using our model for the experimental parameters of Boffetta et al.~\cite{Boffetta:2007EPL} on a log-log scale.  The red, green, and blue curves represent the $\alpha=0.037, 0.059, 0.069$ experimental runs respectively (see Table 1). The power law function $k^{-a}$ (dashed curves) does not fit with the energy spectrum. }
 \label{fig:Boffetta_powerlaw}
 \end{figure}
  \begin{figure}[htbp]
 \begin{center}
 \includegraphics[scale=0.32]{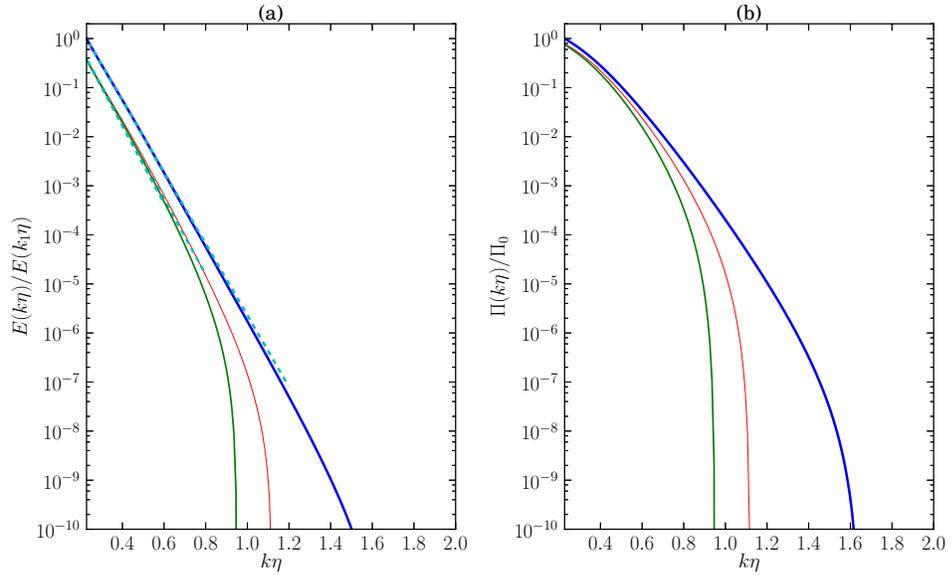}
 \end{center}
 \caption{  (a) Plots of energy spectra for the experimental parameters of Boffetta et al.~\cite{Boffetta:2007EPL} with  $\alpha'=0.037$ (red thin curve), $0.059$ (green thicker curve), and $0.069$ (blue thickest curve) on a semi-log scale.  See Table 1 for more details.  Exponential functions $\exp(-b(k\eta-k_1\eta))$ (dashed cyan lines) with $b=16.7$, 18.1, and 16.8 respectively are the best fit curves for the spectra in the non-viscous range.   (b) The corresponding enstrophy flux for the aforementioned experimental cases.}
 \label{fig:Bofetta_exponential}
 \end{figure}

For the three aforementioned experimental sets of Boffetta et al.~\cite{Boffetta:2007EPL}, the nondimensional parameters are $\alpha'=0.046, 0.086, 0.107$ and $\operatorname{Re}_\alpha = U /(\alpha L) =  0.71,0.45,0.22$ respectively, indicating that the nonlinear term is somewhat weaker than the Ekman friction.  We also estimate the maximum enstrophy flux $\Pi_0$ as $ \omega_\mathrm{rms}^2/T \approx \omega_\mathrm{rms}^3 $.  Using the aforementioned parameters we compute the energy spectra and the normalized enstrophy fluxes for these experimental sets using our model, and compare the results.   In Fig.~\ref{fig:Boffetta_powerlaw} we plot the energy spectra and their best power-law fits.   The figure illustrates that a power law is not a good fit for the energy spectra predicted by the model.   Therefore, we attempt to fit the energy spectra with an exponential function  $\exp(-b (k\eta-k_1\eta))$ for $k>k_1$.   In Figure~\ref{fig:Bofetta_exponential}  we plot the normalized energy spectra  $E(k\eta)/E(k_1\eta)$ and normalized enstrophy flux on a semi-log scale.  The exponential functions (the dashed lines) fit well with the energy spectra in the non-viscous range, and the fitting parameters $b$ for the three experimental cases are 16.7, 18.1, and 16.8 respectively.    Thus an exponential function may describe the energy spectrum of the Boffetta et al.~better than a power-law function, consistent with the observation that Ekman friction dominates the nonlinear term in these cases ($\operatorname{Re}_\alpha < 1$).     Also, the approximate enstrophy flux computed using Eq.~(\ref{eq:Pi_approximate}) does not match with the model flux [Eq.~(\ref{eq:Pi_analytic})] due to absence of an inertial range, which is due the dominance of Ekman friction over the nonlinear term.  A cautionary remark however is in order: our estimate of $\Pi_0$, $Re$ using the experimental parameters of Boffetta et al.'s~\cite{Boffetta:2007EPL} have significant errors. Hence the model predictions are somewhat uncertain, and they need to be verified carefully using more refined experiments and numerical simulations.

In summary, we showed that Ekman friction strongly affects the enstrophy flux in two dimensional turbulence.  We derived an expression for the variable enstrophy flux in terms of Ekman friction parameter $\alpha$.  As a result, the kinetic energy spectrum is steeper than $k^{-3}$.  For large Reynolds number flows with $ \operatorname{Re}_\alpha = U/(\alpha L) > 1$, we expect a power law behaviour for the energy spectrum.  This is due to the dominance of the nonlinear term over Ekman friction.  However, Ekman friction is stronger than the nonlinear term for flows with $ \operatorname{Re}_\alpha < 1$, hence an exponential energy spectrum may be expected for such flows.  

The method presented here is quite general and  could be applied to other situations as well.  The liquid metal flows  in the presence of a strong magnetic field under the quasistatic approximation~\cite{Knaepen:2008ARFM} has a dissipation term similar to the Ekman friction.   Adaption of the present procedure to the liquid metal flow would provide valuable insights into its energy spectrum and dynamics.   Another important direction for future work could be verification of our model using direct numerical simulation.  

I thank Mani Chandra for Python related help and useful comments.  I also thank Sagar Chakraborty, Prasad Perlekar, Stefano Musacchio, and Anupam Gupta for the discussions.

\bibliographystyle{eplbib}

\pagebreak
 \begin{center}
 {\bf List of Tables}
  \end{center}

 \begin{table}[ht]
\caption{Table depicting our model calculation for the Reynolds number $\operatorname{Re}=10^6$ and $\alpha'=0, 0.3, 0.5$  (first three rows), and for the experimental parameters of Boffetta et al.'s~\cite{Boffetta:2007EPL} (next three rows).   Here $\alpha$ is the coefficient of the Ekman friction,  $u_\mathrm{rms}$ is the rms velocity, $\omega_\mathrm{rms}$ is the rms   vorticity,   $\Pi_0$  is the maximum value of enstrophy flux,  $\alpha^\prime = \alpha C/(3\Pi_0^{1/3})$ with $C=1.4$,  $\operatorname{Re}_\alpha= U/(\alpha L)$ is the ratio of the nonlinear term and the Ekman friction term,  $a,a_B$ are the spectral exponents computed using our model and by Boffetta et al. respectively,  $b$ is the coefficient of $E(k) = \exp(-b(k\eta-k_1\eta))$, and $D_\alpha/D_\nu$ is the ratio of the dissipation rates by Ekman friction and the viscous force.  For Boffetta et al.'s experimental parameters~\cite{Boffetta:2007EPL}, we use $Re_\alpha = u_{\mathrm{rms}}/(\alpha L)$ with $L=50$ cm, and $\Pi_0 = \omega_{\mathrm{rms}}^3$. }
\vspace{0.5cm}
\begin{tabular}{ c c c c c c c c c c c} \hline
$\alpha$ &  $u_{\mathrm{rms}}$ &  $\omega_{\mathrm{rms}}$  & $\Pi_0$ & $\alpha^\prime$ & $Re$ & $Re_\alpha$ & $a$  & $a_B$ &  $b$ & $D_\alpha/D_\nu$ \\
s$^{-1}$ & cm/s  & s$^{-1}$ & s$^{-3}$ & -  & - & - & - & - & - & - \\  \hline
 - 		& -  &  - 	&  - & 0 	& $10^6$ 	& $\infty$	& 3  &  -  & - & 0 \\
  - 		&  - &  - & - & 0.3 	& $10^6$ 	&  $\approx 3$ 	& 3.8  &  -  & - & 22 \\
   - 		& -  &   - 	&  - & 0.5 	& $10^6$ 	&  $\approx 2$	& 4.2  &  -  & - & 100 \\ \hline
0.037	& 1.32 &    0.75  & 0.42  	& 0.046 & 6600 	&  0.71 	& - & 3.5 & 16.7 & 0.091 \\
0.059	& 1.33 &   0.64  & 0.26	& 0.086 & 6650	&  0.45	& -  & 3.8 & 18.1 & 0.18 \\
0.069	& 0.79 &   0.60  & 0.21	& 0.107 & 3950	&  0.22	&-  & 4.0 &  16.8   & 0.16 \\ \hline
\end{tabular}

\label{Table1}
\end{table}

\end{document}